\documentclass[sigconf]{acmart}
\acmConference[MSR 2022]{MSR '22: Proceedings of the 19th International Conference on Mining Software Repositories}{May 23–24, 2022}{Pittsburgh, PA, USA}

\usepackage{caption}
\usepackage{subcaption}
\usepackage{float}
\usepackage{graphicx}
\usepackage[flushleft]{threeparttable}
\usepackage{enumitem}
\usepackage{balance}
\usepackage{xspace}
\usepackage{multirow}
\newcommand{\Fig}{Figure\xspace}
\newcommand{\Tab}{Table\xspace}
\newcommand{\Sec}{Section\xspace}

\newcommand\base{Base\xspace}
\newcommand\baseWithFcg{Base$_{+fcg}$\xspace}
\newcommand\libdx{LibDX\xspace}
\newcommand\bsfinder{B2SFinder\xspace}
\newcommand\bsfinderWithFcg{B2SFinder$_{+fcg}$\xspace}

\newcommand\oursnospace{LibDB}
\newcommand\ours{LibDB\xspace}
\newcommand{\oursBaseWithoutFCG}{\oursnospace$_{base-fcg}$\xspace}
\newcommand{\oursBase}{\oursnospace$_{base}$\xspace}
\newcommand{\oursFRWithoutFCG}{\oursnospace$_{fr-fcg}$\xspace}
\newcommand{\oursFR}{\oursnospace$_{fr}$\xspace}

\newcommand{\oursWithoutFcg}{\oursnospace$_{-fcg}$\xspace}

\newcommand\baseWithFR{Base$_{+fr}$\xspace}
\newcommand\bsfinderWithFR{B2SFinder$_{+fr}$\xspace}
\newcommand{\oursWithoutFR}{\oursnospace$_{-fr}$\xspace}

\newcommand{\datasetOne}{Cross-5C Dataset\xspace}
\newcommand{\datasetTwo}{TPLBinary Dataset\xspace}
\newcommand{\datasetThree}{FedoraLib Database\xspace}

\newcommand{\saveSpaceFig}{\vspace{-6pt}}
\newcommand{\saveSpaceText}{\vspace{4.5pt}}

\AtBeginDocument{%
  \providecommand\BibTeX{{%
    \normalfont B\kern-0.5em{\scshape i\kern-0.25em b}\kern-0.8em\TeX}}}

\copyrightyear{2022}
\acmYear{2022}
\setcopyright{acmlicensed}\acmConference[MSR '22]{19th International Conference on Mining Software Repositories}{May 23--24, 2022}{Pittsburgh, PA, USA}
\acmBooktitle{19th International Conference on Mining Software Repositories (MSR '22), May 23--24, 2022, Pittsburgh, PA, USA}
\acmPrice{15.00}
\acmDOI{10.1145/3524842.3528442}
\acmISBN{978-1-4503-9303-4/22/05}

\begin{document}

\title{LibDB: An Effective and Efficient Framework for Detecting Third-Party Libraries in Binaries}

\author{Wei Tang}
\email{tang-w17@mails.tsinghua.edu.cn}
\authornote{Work is done during internship at Microsoft Research Asia.}
\affiliation{
  \institution{School of Software, Tsinghua University}
  \city{Beijing}
  \country{China}
}
\author{Yanlin Wang}
\authornote{Corresponding author.}
\email{yanlwang@microsoft.com}
\affiliation{%
  \institution{Microsoft Research}
  \city{Beijing}
  \country{China}
}
\author{Hongyu Zhang}
\email{hongyu.zhang@newcastle.edu.au}
\affiliation{%
  \institution{The University of Newcastle}
  \city{Newcastle}
  \country{Australia}
}
\author{Shi Han}
\email{shihan@microsoft.com}
\affiliation{%
  \institution{Microsoft Research}
  \city{Beijing}
  \country{China}
}
\author{Ping Luo}
\email{luop@mail.tsinghua.edu.cn}
\affiliation{%
  \institution{School of Software, Tsinghua University}
  \city{Beijing}
  \country{China}
}
\author{Dongmei Zhang}
\email{dongmeiz@microsoft.com}
\affiliation{%
  \institution{Microsoft Research}
  \city{Beijing}
  \country{China}
}

\renewcommand{\shortauthors}{Tang, et al.}

\begin{abstract} 
Third-party libraries (TPLs) are reused frequently in software applications for reducing development cost. However, they could introduce security risks as well. Many TPL detection methods have been proposed to detect TPL reuse in Android bytecode or in source code. This paper focuses on detecting TPL reuse in binary code, which is a more challenging task. For a detection target in binary form, libraries may be compiled and linked to separate dynamic-link files or built into a fused binary that contains multiple libraries and project-specific code. This could result in fewer available code features and lower the effectiveness of feature engineering. 
In this paper, we propose a binary TPL reuse detection framework, LibDB, which can effectively and efficiently detect imported TPLs even in stripped and fused binaries. In addition to the basic and coarse-grained features (string literals and exported function names), LibDB utilizes function contents as a new type of feature. It embeds all functions in a binary file to low-dimensional representations with a trained neural network. 
It further adopts a function call graph-based comparison method to improve the accuracy of the detection. LibDB is able to support version identification of TPLs contained in the detection target, which is not considered by existing detection methods. 
To evaluate the performance of LibDB, we construct three datasets for binary-based TPL reuse detection.
Our experimental results show that LibDB is more accurate and efficient than state-of-the-art tools on the binary TPL detection task and the version identification task. 
Our datasets and source code used in this work are anonymously available at \textbf{\url{https://github.com/DeepSoftwareAnalytics/LibDB}}.
\end{abstract}



\maketitle

\section{Introduction}

Third-party libraries (TPLs) are important components of modern software systems. They are reused frequently during software development~\cite{zhang2019libid, Wang2015WuKongAS}. Open-source repository platforms and package management systems are the major sources of third-party libraries. However, security issues of the third-party code continue to arise. Vulnerabilities in well-known third-party libraries, such as the \emph{Heartbleed}~\cite{durumeric2014matter} bug, could bring security threats to millions of devices. In addition, non-compliant reuse, which is a violation of legal software licenses, could lead to costly commercial disputes. Unfortunately, many developers do not pay sufficient attention to the code that is imported from third-party libraries.

To overcome the emerging threats caused by the reuse of third-party libraries, a great number of research works~\cite{Jang2012ReDeBugFU, Kim2017VUDDYAS, Duan2017Identifying, li2017libd, zhang2019libid, zhan2020automated} and commercial products~\cite{blackduck, whitesource} have been proposed. They detect TPL reuse based on a database of libraries, identify their versions that may contain potential vulnerabilities, and manage license violation risks. Most of them handle detection targets that are in the forms of source code ~\cite{Jang2012ReDeBugFU, Kim2017VUDDYAS, woo2021centris} or Java bytecode ~\cite{zhan2020automated}. 
TPL detection for binaries compiled from C/C++ sources is also important since many projects (especially the ones that require high efficiency) are written in C/C++. However, only a few previous works studied this problem~\cite{Zhan2021ResearchOT}. It is urgent to fill the gap of TPL detection in binaries. 
Compared to source code or bytecode forms, it is more difficult to detect TPLs for binaries as features such as variable names and function names are stripped after compilation. 

Most existing approaches for TPL detection in binaries, including BAT~\cite{Hemel2011FindingSL}, OSSPolice~\cite{Duan2017Identifying} and \libdx~\cite{libdx}, use two types of basic features, i.e., string literals and exported function names. Only \bsfinder~\cite{b2sfinder} proposes to use extra features including \verb|integer constants|, \verb|switch/else|, and \verb|if/else| to detect commercial off-the-shelf (COTS) software that may be stripped of strings and exported function names. However, \bsfinder still mainly relies on the basic features for detection.
Basic features are highly efficient to search by indexing as key-value pairs, however, they have many disadvantages. As the code content is ignored and basic features may exist cross libraries, they are not sufficiently unique to represent a library in a large-scale database. They are also too coarse-grained to identify specific versions, since basic features may not change cross versions. Besides, not all basic features are necessarily retained in binaries after compilation with different settings.

A detailed discussion of these limitations can be found in \Sec~\ref{motivation}. 

To address the limitations of basic features used in existing work, function-level features can be utilized, as they contain more fine-grained information that can represent functions and their call relationships in binaries. Specifically, in our work, we incorporate function vector features to improve the recall of TPL detection, because they can be fully/partially preserved in binaries. We also design a novel filter module to filter away wrongly reported libraries to further improve the precision of TPL  detection. 

In this paper, we propose a framework \textbf{\ours} for accurately and efficiently detecting third-party libraries in binaries using additional function-level features and function call graphs (FCGs). We adopt a binary-to-binary matching method and build the local TPL feature database containing features extracted from TPLs in binary form. 
In addition to the basic features (i.e., string literals and exported function names), we embed all functions in a binary to low-dimensional representation vectors via a graph embedding network, connect all functions through dependency, and obtain FCGs to represent the binary.
Against the local feature database, 
\ours uses two channels (the basic feature channel and the function vector channel) to quickly detect candidate TPLs that may be contained in the detection target. 
Then, it compares each candidate to the detection target using the FCG representation.
Candidates from two channels, after using FCG filter, are integrated as the final detection results.
Unlike existing binary-oriented TPL detection methods, \ours can further provide version identification of the TPLs contained in the detection target.

In summary, our contributions are as follows:
\begin{itemize}
\item We propose a novel framework \ours that utilizes function contents to detect binary TPL reuse instead of relying on basic features only.
\item We design a comparison algorithm based on FCGs to calculate the similarity between two binary files. The algorithm greatly improves the precision of both \ours and existing approaches. 
\item We use fined-grained function features to better identify the versions of the reused libraries, which is not supported by previous binary TPL detection methods.
\item We have conducted extensive experiments to illustrate the effectiveness and efficiency of \ours compared to state-of-the-art approaches.
\end{itemize}

\begin{figure*}[t]
    \centering
    \includegraphics[width=0.85\textwidth]{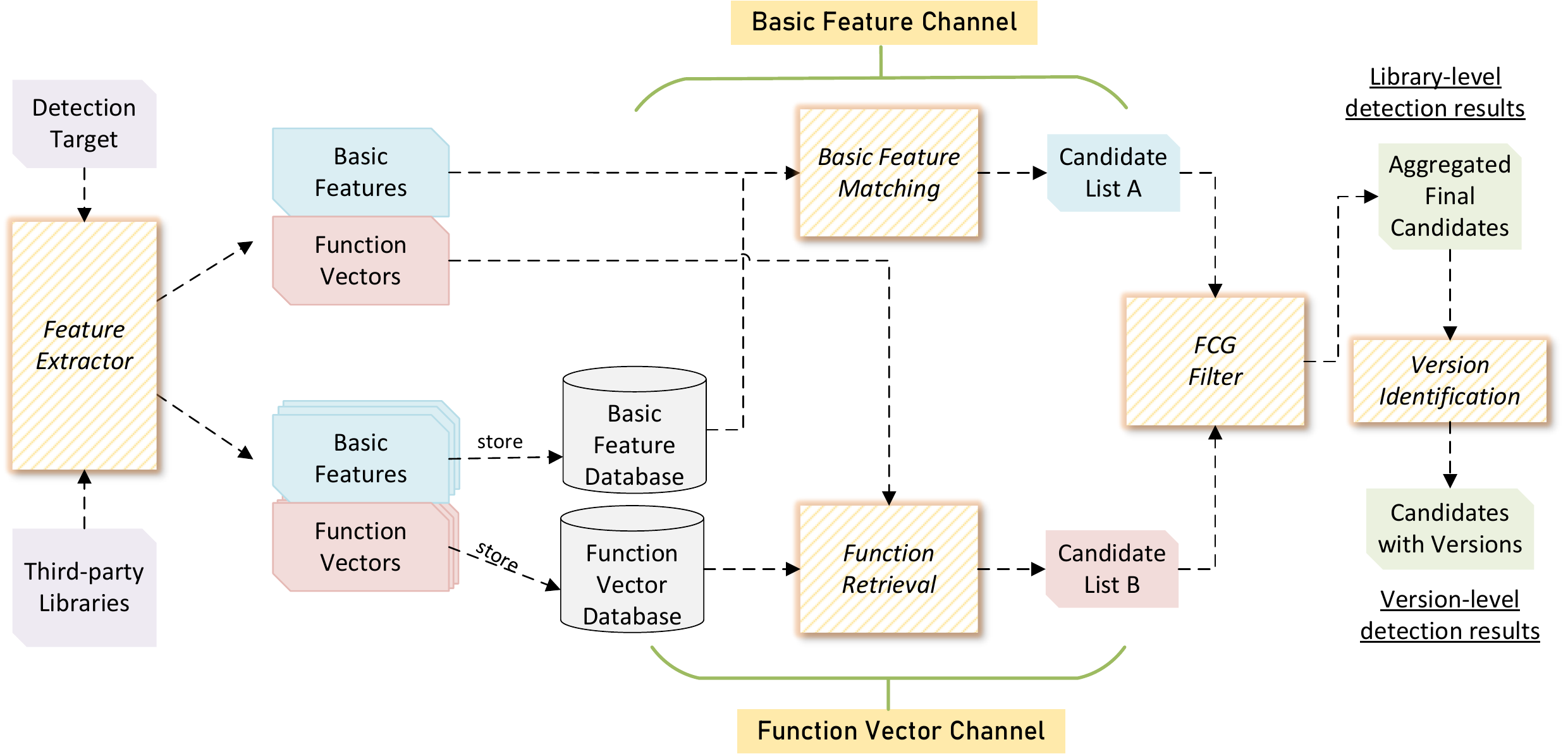}
    \vspace{-12pt}
    \caption{Overview of \ours.}
    \vspace{-5pt}
    \label{fig:architecture}
\end{figure*}

\section{Background and Motivation}\label{motivation}
\subsection{Existing Work on TPL Detection for Binaries}
Existing techniques~\cite{Duan2017Identifying, b2sfinder, libdx} for third-party library detection in binaries essentially have a similar detection scheme.
They construct two sets of features from a detection target ($BIN$) and a library ($LIB_{UNIT}$). The task to detect library usage is to calculate a similarity score by comparing these two feature sets ($\frac{\mid BIN\cap LIB_{UNIT}\mid}{\mid LIB_{UNIT}\mid}$). Existing tools divide a library into multiple comparison units and a set of features is extracted from each unit. A similarity score is assigned to each unit and a positive unit means a library reuse. Weighting functions may be used to assign different scores to features. BAT~\cite{Hemel2011FindingSL} treats the entire library repository as a unit and extract string literals as features. It considers longer strings to have more uniqueness and assigns them larger weights. OSSPolice~\cite{Duan2017Identifying} indexes a repository by its structural layout (i.e., a tree of files and directories) and takes each node (a directory) as a comparison unit to calculate a score. \libdx~\cite{libdx} takes each binary in a library as a unit.
It only considers the features of continuous matching, which means that it may ignore a large number of features, resulting in a low recall rate. 
Both OSSPolice and LibDX use TF-IDF to calculate a weighted score. 

A state-of-the-art tool, B2SFinder~\cite{b2sfinder}, separates a library to multiple groups by parsing compilation commands, where all sources are organized as compilation dependency graphs and each graph is a comparison unit for similarity calculation. It assigns larger scores to special strings such as web links, function names, etc. B2SFinder introduces additional features including integer constants, constants in conditional statements (``\verb|switch/case|'' and ``\verb|if/else|''), and constant arrays. In essence, B2SFinder still heavily depends on basic features. It considers the order of features to construct array data and the structure of conditional statements. Besides, new types of features require B2SFinder to adopt an one-to-one comparison method for all libraries in database, since they cannot be implemented by searching. It limits the capability of B2SFinder when the database grows. In summary, all existing techniques heavily rely on basic features (string literals and exported function names).

\subsection{Limitations of Existing Work}
Existing works that compare two sets of basic features have the following deficiencies when faced with fused binaries:

\saveSpaceText
\noindent\textbf{Low precision when detecting against a large-scale database.} Some features are popular and included in a large number of libraries. As the size of the library database increases, the number of possible occurrences of each feature increases, therefore the uniqueness and effectiveness of basic features decrease. Since multiple libraries are compiled into a fused binary, more popular features will be accumulated. When detecting against a larger database, more false positives could be reported~\cite{woo2021centris}. 

\saveSpaceText
\noindent\textbf{Low recall when few basic features exists.} 
We observed that nearly all string literals are printout messages, such as log information including debug, warnings and errors. Developers can select the information they want and even remove all printouts by setting compilation macros. The compilation settings (i.e., ``\verb|FLAGS|'') determine the code to be compiled in the compilation phase. For example, function ``\verb|PNG_DEBUG|'' in \texttt{LibPNG} prints information when debugging. It contains message literals as arguments, however, they would not be compiled into the released files by undefining ``\verb|PNG_DEBUG|'' in compilation parameters. In \texttt{LibPNG}, more than 95\% of strings are printouts and most can be removed by macro settings. Besides, Quach et al. ~\cite{Quach2018DebloatingST} shows that only 5\% of \texttt{libc} functions are used on average by Ubuntu programs. Binary debloating techniques~\cite{Quach2018DebloatingST, Agadakos2019NibblerDB} directly eliminate unneeded functions to improve software quality, meaning that only a small portion of functions from a source code file might be preserved. If the features are extracted at the file level, binary debloating would lead to a low similarity of the file, causing false negatives for approaches such as OSSPolice, B2SFinder, and LibDX, since they extract features and generate signatures at the file level. 

\saveSpaceText
\noindent\textbf{Coarse-grained representation for version identification.} TPLs evolve over time, with known vulnerabilities being patched and new (potentially vulnerable) code being added.
Therefore, vulnerabilities are usually present in some successive versions. Each new version only makes minor changes to the previous one and different versions of a library are similar, especially for adjacent versions. The slightly changed code (e.g., a patch) has little effect on basic features. Therefore, previous approaches using coarse-grained model with basic features might not capture these subtle changes between versions. Only OSSPolice shows the results for version identification using unique strings to determine the best version, such as ``\verb|1.2.6|''. However, the library information may not be stored in fused binaries. For example, the binary ``\verb|libmain.so|'' in ``\verb|Gomoku|''\footnote{\url{https://f-droid.org/en/packages/com.traffar.gomoku/}} reuses the library ``\verb|AGG|'', but there is no information about ``\verb|AGG|'' in the version string. On the other hand, information about other libraries may provide noisy version strings. OSSPolice visioned that it can be empowered by fine-grained function-level features to improve the precision of version identification.

Our work addresses the limitations of existing work. To deal with the low precision problem, we design a FCG filter to filter out false positives. To deal with the low recall problem, in addition to the basic features, we identify  function-level features to retrieve more results from the library database. Our fine-grained function features and FCG filter together can distinguish more correct versions than the coarse-grained basic features.

\section{Proposed Approach}\label{design}

\subsection{Overview of \ours}
\Fig~\ref{fig:architecture} provides an overview of \ours. \ours contains five {components}: Feature Extractor (\Sec~\ref{sec:fe}), Basic Feature Matching (\Sec~\ref{sec:bfm}), Function Retrieval (\Sec~\ref{sec:fr}), Function-call Graph (FCG) Filter (\Sec~\ref{FCG comparison}), and Version Identification (\Sec~\ref{sec:vi}). \ours relies on the two {feature databases}, namely basic feature database and function vector database, which will be applied in two {channels}: basic feature channel and function vector channel, during detection.
In the following, we briefly introduce each key component of \ours to help readers acquire a high-level understanding of the workflow of \ours before we step into its detailed designs:

\saveSpaceText
\noindent\textbf{Feature Extractor.} Both TPLs and detection targets are in binary form and they share the feature extractor module. Binary inputs are disassembled and parsed by the feature extractor module. In the feature extractor, string literals and exported function names are extracted as basic features. Functions are embedded to representation vectors using neural networks and similar functions are mapped to representation vectors that have high cosine similarity scores. 

\saveSpaceText
\noindent\textbf{Feature Databases.} The output of the feature extractor contains two types of features: basic features and functions, which constitute the basic feature database and the function vector database, respectively. When building the local TPL database, each binary in a downloaded package is regarded as a comparison unit. Similarity will be calculated between each comparison unit and the detection target. In two feature databases, using inverted index, each feature is mapped to the comparison unit where the feature is extracted.

\saveSpaceText
\noindent\textbf{Channels.}
When matching features of detection targets, there are two channels corresponding to the basic features and function features. 
In the basic feature channel, basic features are applied to search directly in the basic feature database to obtain initial candidates rapidly. 
In the function vector channel, we retrieve the top-$K$ nearest neighbors in the function vector database and locate related comparison units as initial candidates. 

\saveSpaceText
\noindent\textbf{FCG Filter.}
Due to the lack of uniqueness of features against a large-scale database, there might be many false positives in the initial candidates output from the channels. 
We compare each candidate with the detection target using function-call graphs (FCG) in the FCG filter. 
FCG filter uses the same matching methods as the two channels but set different thresholds. The reasons and more descriptions are provided in \Sec~\ref{FCG comparison}. After using FCG to filter candidates, we get the final candidates. By locating libraries where the candidates are packaged, we can report library-level detection results. 

\saveSpaceText
\noindent\textbf{Version Identification.}
\ours further adopts a version ranking method to each positive library output by the FCG filter, then it pinpoints the rank \#1 candidate as the version-level detection result.

\subsection{Feature Extractor}
\label{sec:fe}

\begin{figure}[t]
    \centering
    \includegraphics[width=0.7\columnwidth]{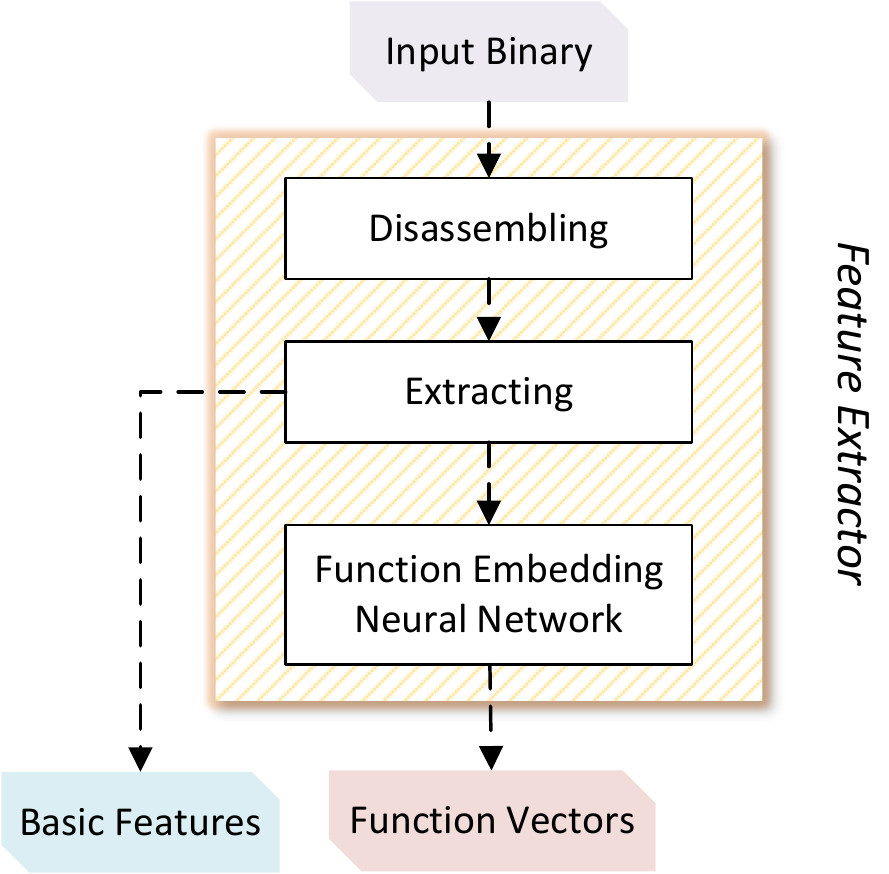}
    \saveSpaceFig
    \caption{Feature extractor module.}
    \saveSpaceFig
    \label{fig:feature_extractor}
\end{figure}

\Fig~\ref{fig:feature_extractor} illustrates the design of the feature extractor. Feature extractor extracts basic features and function vectors to construct fingerprints via three steps: disassembling binaries, extracting features, and function embedding. 

\subsubsection{Disassembling and Extracting}

We build a Ghidra-based module to disassemble binaries, parse the assembly code, and extract features we need. Ghidra\footnote{\url{https://ghidra-sre.org}} is an open source reverse engineering framework developed by NSA (National Security Agency of the United States).
It allows modular removal to accelerate the reverse process. After binaries are transformed to assembly code, we extract three parts: string literals, exported function names, and function features. 

String literals and exported function names are basic features. String literals can be easily extracted from the data segment (\verb|.rodata|, \verb|.data| and \verb|cstring|). We can obtain exported function names by reading their symbols in binaries. 
We further extract control flow graphs (CFGs) of functions as function features, which are used to generate function vectors in the subsequent embedding step.

\subsubsection{Function embedding} \label{sec:femb}
Real-world applications have a wide variety of compilation conditions. 
Different compilation conditions may change binary code that are compiled from the same source code, resulting in difficulties in binary code matching.

Considering TPL reuse in real-world applications, we have summarized the following five main scenarios (5C scenarios) that could introduce changes to the binary code:
\begin{itemize}
\item \underline{\textit{Cross-operating-system.}} Open-source libraries may have different macro definitions depending on the target operating systems. For example, in the TPL \texttt{LibPNG}, there are conditional statements like ``\verb|#if defined(WIN32)|'' and different ``\verb|Makefile|'', which change the contents of compiled binaries on different operating systems. Besides, the format of binary files depends on the system they run on. Most common formats are PE (on Microsoft Windows), ELF (on Linux), and Mach-O (on MacOS).

\item \underline{\textit{Cross-architecture.}} Content in compiled binaries varies with CPU architectures. For instance, there are several folders in \texttt{LibPNG} such as ``\verb|arm|'', ``\verb|intel|'', ``\verb|mips|'' that contain unique optimized code for corresponding CPU architectures. Macro ``\verb|PNG_ARM_NEON_IMPLEMENTATION|'' controls whether to use certain functions. It is also possible to set its value in compiler settings. In this case, the final constant features contained in binary files cannot be determined from the original code repository without compilation. 

\item \underline{\textit{Cross-compiler.}} In the open-source ecosystem, developers usually use GCC~\cite{gcc} as the main compiler. However, LLVM~\cite{llvm} has become more popular in recent years because of its excellent performance. Different compilers make differences when reusing TPLs.

\item \underline{\textit{Cross-compiler-version.}} Compilers, like other software, evolve over time. After a library is released, developers may use another version of compiler to recompile their applications. Different versions of compilers also have impact on the binary code.

\item \underline{\textit{Cross-optimization-level.}} Optimization is one of the challenges for binary code similarity detection~\cite{genius}. Different optimization levels (O0-O3) have significant impact on the binary code. Developers can easily set the optimization level while compiling the reused libraries, resulting in differences in the binary.
\end{itemize}

Then, we construct a large dataset, the \datasetOne, to train the neural network model and calculate the similarity of binaries in 5C scenarios to cope with TPL reuses in real world. More details about the \datasetOne are provided in \Sec~\ref{dataset}. From related works introduced in \Sec~\ref{related work}, we collect 13 open-source projects and compile each one using various compilation settings. For each scenario, we take several different values as possible conditions. 
For the cross-operating-system scenario, we consider three common platforms, Linux, MacOS and Windows. 
For the cross-architecture scenario, we select ARM and x86 as possible architectures. 
For the cross-compiler scenario, we choose two of the most well-known compilers, GCC and LLVM. They have many released versions up to now. We sample versions evenly through the whole version history. All versions are GCC: 4.8, 5.4, 6, 7.1, 8.1, 9, and LLVM: 3.5, 6, 12.
For the cross-optimization-level scenario, we use all four levels (O0-O3) as available conditions. 
We obtain all possible combinations in 5C scenarios, finally we have 120 different kinds of compilation setting combinations.

Binary code similarity detection faces the challenge of 5C scenarios. However, it should be noted that LibDX is neither a system to find vulnerable function in binary code nor does it aim to report the similarity of two functions. We retrieve similar function pairs as features to find reused libraries using existing techniques for detecting similar function in binary code. A relatively large amount of false positive pairs would be retrieved against our large-scale database that contains 18.6 million functions. Accordingly, we designed a FCG-based filter to eliminate false pairs. More details are described in the Section~\ref{FCG Filter}.

For similar function retrieval in Channel 2, CFG based comparison is accurate but cannot handle an enormous amount of TPLs in database. Efficiently searching to retrieve similar functions is crucial, since we care about library-level detection instead of function-level or snippet-level detection. In recent years, researchers have used neural networks to map assembly code of functions to representation vectors~\cite{Xu2017NeuralNG, ding2019asm2vec, Gao2018VulSeekerAS}. Similar functions that are compiled from the same source code are mapped to vectors within a smaller distance than dissimilar functions. It is highly efficient to retrieve nearest neighbors as the detection result via vector similarity search.

\begin{figure}[t]
    \centering
    \includegraphics[width=0.7\columnwidth]{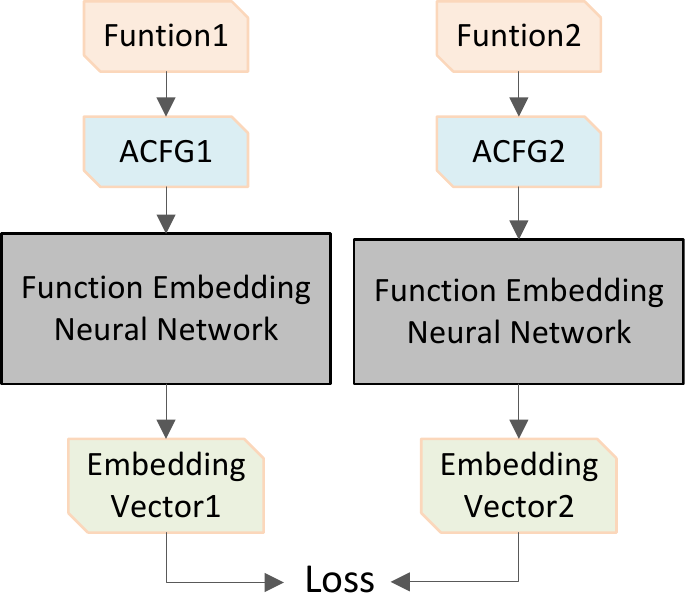}
    \saveSpaceFig
    \caption{Embedding module.}
    \saveSpaceFig
    \label{fig:embedding_module}
\end{figure}

Compared to other models such as ASM2VEC~\cite{ding2019asm2vec} that is mainly designed for a single assembly code language and not applicable for comparison across architectures, Gemini~\cite{Xu2017NeuralNG} is applicable to all 5C scenarios in the field of binary comparison. We use the Gemini model to embed each function to a vector, as depicted in \Fig~\ref{fig:embedding_module}. Firstly, each function is transformed into an attributed control flow graph (ACFG) in which each block is represented using manually selected 7 types of features. Following the related work~\cite{Yu2020CodeCMRCR, Xu2017NeuralNG}, we ignore functions with fewer than 5 basic blocks. We then adopt Structure2vec~\cite{dai2016discriminative} as the graph embedding network. It converts a graph (i.e., ACFG in \ours) to a representation vector. A Siamese architecture~\cite{bromley1993signature} that contains one shared-parameter function embedding network (i.e., Structure2vec) is applied as shown in \Fig~\ref{fig:embedding_module}. More technical details about Gemini can be found in its original paper [30].

We use the contrastive loss in Equation~\ref{eq:loss} for the optimization of the function embedding module. Optimizing Equation~\ref{eq:loss} leads similar function pairs (i.e., functions compiled from the same source code under different compilation conditions) to have cosine similarity close to 1, and dissimilar pairs (i.e., functions compiled from different source code) to have cosine similarity close to -1. In Equation~\ref{eq:loss}, $Y_{i,j}$ represents the label of a function pair $f_i$ and $f_j$ and $\mathcal{F}$ is the set of functions. For each function, we randomly sample one similar function compiled from the same source code under different compilation conditions, and one dissimilar function compiled from different source code. If two functions are similar, their label is 1, otherwise -1. $S(f_i, f_j)$ is the cosine similarity of the two representation vectors of $f_i$ and $f_j$. This way, we can efficiently search function representation vectors of detection targets against a large-scale database and retrieve similar functions according to the cosine similarity.
\begin{equation}
\label{eq:loss}
\mathcal{L} = \sum_{f_i,\,f_j \in \mathcal{F}} \frac{1}{2}\big(1+Y_{i,j}\big)\big(1-S(f_i, f_j)\big)^2+\frac{1}{2}\big(1-Y_{i,j}\big)\big(1+S(f_i, f_j)\big)^2. 
\end{equation}

\subsection{Basic Feature Matching}
\label{sec:bfm}

Basic feature channel searches basic features in the basic feature database and obtain a list of initial candidates (Candidate List A in \Fig~\ref{fig:architecture}). For efficiency, we optimize the matching process using inverted index where string literals and exported function names are keys and comparison units the key exists in as values. We search basic features in database and get a list of corresponding comparison units as candidates. Each candidate has a set of common features that exist in both the comparison unit and detection target. We directly adopt the rules and thresholds used in \bsfinder~\cite{b2sfinder} to filter candidates. A candidate is positive when it meets any of the following conditions: (1) the proportion of common strings against comparison unit is larger than 0.5; (2) the sum of weights is larger than 100 and the proportion of weights is larger than 0.1; (3) the number of common exported function names is larger than 20.
More details can be found in \bsfinder repository.\footnote{https://github.com/1dayto0day/B2SFinder} Positive units are initial candidates and are sent to the FCG filter as inputs.

\subsection{Function Retrieval}
\label{sec:fr}

In addition to basic feature matching, in the function vector channel, \ours can retrieve candidates using functions. This capability is very important especially when there are few or no basic features. We have collected a large-scale TPL database, namely the \datasetThree, which contains around 1,000 libraries with 25,000 versions from Fedora mirrors.\footnote{\url{https://admin.fedoraproject.org/mirrormanager/}} 
More details about the \datasetThree are provided in Section~\ref{dataset}. 
All binaries as comparison units in TPLs are processed by the feature extractor introduced in \Sec\ref{sec:fe}. In total, 18.6 million functions are embedded to representation vectors and stored in the function vector database. 

In order to detect the potential reuses of TPLs in detection targets, all functions of a detection target are extracted and converted to representation vectors using Structure2vec. Since all similar functions are mapped to closer vectors, we are able to search for similar functions using the nearest neighbor algorithm. We retrieve the top-$K$ most similar functions for each function in the detection target and locate corresponding comparison units as initial candidates (Candidate List B in \Fig~\ref{fig:architecture}). A larger K means that more functions are retrieved as similar functions. Further, more comparison units are obtained.
On the contrary, a smaller K means fewer candidates but they are more likely to be similar to the function in the detection target. We implement the function retrieval module based on an efficient vector searching engine Milvus.\footnote{\url{https://milvus.io}} It supports using GPU and creates an index to speed up the searching phase. We choose the inner product distance between two embeddings as the similarity metric to search, since inner product can be easily converted to cosine similarity, which we use in the function embedding module (\Sec~\ref{sec:femb}).

If the target has $M$ functions, we can get $M*K$ similar functions and locate comparison units where these functions are extracted from. We sort all comparison units according to the number of similar functions they contain. Top-$200$ units will be passed to the subsequent FCG filter.

\subsection{FCG Filter}\label{FCG Filter}
\label{FCG comparison}

The candidate list obtained from basic feature matching and function retrieval may contain lots of false positives. In order to filter out them, we propose a novel FCG based algorithm to compare binaries with the detection target and filter out those with low similarity. The algorithm contains two steps: function pairing and FCG comparison.

\subsubsection{Function pairing}\label{function pairing}
Given the FCG of a candidate and the FCG of the detection target, function pairing determines the node mapping between them. For each function $f_i$ in the detection target, we match it with the most similar function $f_i^{'}$ that has the largest cosine similarity to $f_i$ among all functions in the candidate. We set a threshold that requires the cosine similarity is larger than 0.8. It is a reasonable value according to our experiments. More details are discussed in \Sec~\ref{accuracy of network model}. Finally, we can obtain a set of similar function pairs $\langle f, f^{'}\rangle$ indicating the similar node pairs between the candidate and the target. Note that for the function vector channel, the function pairing step can be omitted as similar function pairs are already retrieved in the function retrieval module.

\subsubsection{FCG comparison}

\begin{figure}[t]
    \centering
    \begin{subfigure}[b]{\columnwidth}
        \centering
        \includegraphics[width=0.5\columnwidth]{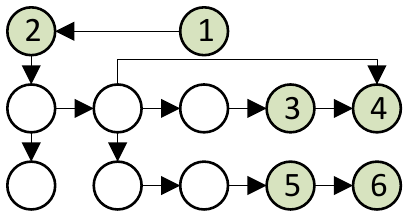}
        \caption{The function call graph of a detection target.}
        \label{fig:func_in_lib}
    \end{subfigure}
    \hfill
    \begin{subfigure}[b]{\columnwidth}
        \centering
        \includegraphics[width=0.7\columnwidth]{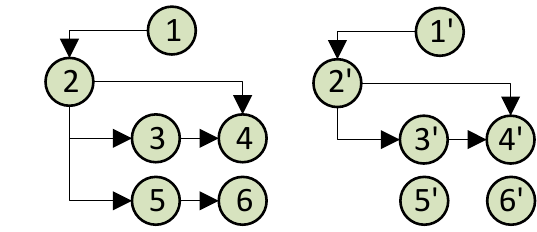}
        \caption{FCG comparison between library and detection target. Left is the mini-FCG for the target. Right is the mini-FCG for the library.}
        \label{fig:fcg_comparison}
    \end{subfigure}
    \vspace{-15pt}
    \caption{The FCG comparison method.}
    \vspace{-10pt}
    \label{fig:fcg_comparison_method}
\end{figure}

After the set of similar function pairs is obtained, we identify correct candidates using FCGs. 
\Fig~\ref{fig:func_in_lib} illustrates the FCG of a detection target where a node represents a function and a directed edge represents function call dependency, with the caller pointing to the callee. Green nodes correspond to functions that exist in the similar function pairs and white nodes correspond to functions that do not exist in similar function pairs. We take the matched nodes (\verb|Node 1-6|) as anchor nodes, since they are key points to show that two FCGs are similar. To reduce the size of graphs, we skip unmatched nodes (i.e., white nodes) to build a mini-FCG using anchor nodes only. In an FCG, unmatched nodes will be removed and dangling edges are connected from upstream nodes to downstream nodes. As shown in \Fig~\ref{fig:fcg_comparison_method}, the corresponding mini-FCG of \Fig~\ref{fig:func_in_lib} is shown on the left of \Fig~\ref{fig:fcg_comparison}. We remove all white nodes. If a matched node transitively depends on another, for example, node2 transitively depends on node3, we add a new edge from node2 to node3 in the mini-FCG. The original dependency relationships are retained. The mini-FCG formed by similar functions in the library is shown on the right of \Fig~\ref{fig:fcg_comparison}.

Our FCG comparison method calculates the number of common edges as the similarity score. Due to the existence of false similar function pairs, we leverage the function dependency to identify true function reuse. In mini-FCGs of the candidate and the target, two edges are common edges when their caller function pair and callee function pair exist in similar function pairs. It can be seen in \Fig~\ref{fig:fcg_comparison} that, the edge from node 1 to node 2 is a common edge, since both node 1 and node 2 form similar function pairs. Node 5 in the left graph is in the set of similar function pairs, but the corresponding matched function (node 5 apostrophe) does not call the function on node 6 apostrophe and node 2 apostrophe does not call node 5 apostrophe, either. Therefore, the similar function pair (node 5 and node 5 apostrophe) cannot contribute to the similarity score.

In the basic feature channel, we set a threshold on the number of common edges. Empirically, a candidate which has less than 3 common edges is recognized as a false candidate. 
Through comparative analysis, we find that similar function pairs in the function vector channel are of higher quality but low quantity compared to the basic feature channel. The reason is that similar function pairs in the function vector channel are retrieved against TPL function vector database. The top-$K$ most similar functions from 18.6 million functions are of high similarity. However, a comparison unit in our TPL database has 90 functions on average. In the basic feature channel, similar function pairs are matched by function pairing from a target to a unit. Not all pairs in the basic feature channel can surely exist in the top-$K$ most similar functions in the function vector channel.

\subsection{Library Report and Version Identification}
\label{sec:vi}

Our TPL database contains four levels of information: library, version, comparison unit, and feature. A library has multiple versions and a version package may contain multiple binaries. Each of them is a comparison unit to be compared to the target file. 
FCG filter outputs merged final candidates from the basic feature channel and the function vector channel. 
\ours reports all related libraries as library-level detection results where the final candidates exist.
Based on those candidates, \ours can further conduct version identification.
It sums all similarity scores of candidates that exist in the same version package. 
The sum of scores is assigned to each version package. 
For a positive library, we report the version with the largest score as the version-level result.

\section{Experimental Design}
\label{experiment design}

\subsection{Datasets}\label{dataset}
We collect 3 datasets for evaluation:

\saveSpaceText
\noindent\textbf{\datasetTwo:} This dataset is used as ground truth for binary TPL detection and version identification. There is no public test data for TPL detection in binaries. The \datasetTwo is established with the reuse relationships of binary software. We contacted authors of related works and got the raw data of OSSPolice~\cite{Duan2017Identifying} and the test data of \libdx~\cite{libdx}. However, B2SFinder~\cite{b2sfinder} can not extract features from the test data of \libdx successfully. Besides, most of the test data in \libdx are separate dynamic link libraries not fused binaries. The raw data of OSSPolice contains a set of Android applications from F-Droid.\footnote{https://f-droid.org} To find the library reuse relationships in fused binaries, we first check their source repositories and get possible reuses like OSSPolice. Then, we analyze compilation commands in repositories and obtain the source code dependency like \bsfinder. 
Finally, we use the reverse engineering tool to disassemble the binary, and find specific evidence of reuse like \libdx. Since all applications are open-source, lots of information like string literals and function names are kept in binaries and can be used as evidence. Reuse relationships without evidence are ignored. Totally, we have 24 Android applications, 112 binaries and 172 reuse relationships. 

\saveSpaceText
\noindent\textbf{\datasetOne:} The \datasetOne contains manually compiled libraries for training and evaluating function comparison model. We collected 13 projects from related work~\cite{genius, Xu2017NeuralNG} covering areas of image processing, database, encryption, etc. They are manually compiled under the 5C scenarios illustrated in \Sec~\ref{sec:fe}. 

Each function is compiled to multiple instances under as many compilation conditions as possible. Sometimes, the compiled code may be the same in different compilation scenarios. In this case, we only keep one instance instead of multiple identical instances. 
Two binary functions are similar when they come from the same source code, even though they may have been changed in different compilation settings. We do not modify any other settings in compilation commands except for the 5C scenarios mentioned in \Sec\ref{sec:fe}. That means function inlining appears in our compilation processes. We use functions in 10 projects as train and validation sets, and the remaining 3 libraries as test set.

\saveSpaceText
\noindent\textbf{\datasetThree:} We collect a large-scale TPL database from a Fedora mirror manager with all historical versions. All mirrors contain 400 thousand packages. It greatly exceeds the size of database used by existing tools. Based on the \datasetThree, we evaluate the recall of function retrieval module at large-scale, accuracy of library reuse detection and version identification. We select libraries that can be found in NVD\footnote{https://nvd.nist.gov} and obtain 997 libraries with about 25,000 versions and more than 200,000 comparison units. It is ten times larger than the TPL database used by B2SFinder~\cite{b2sfinder}. In total, 18.6 million functions are extracted from the \datasetThree.

\subsection{Compared Methods}~\label{sec:baseline_description}
We compare \ours with several related methods and their variants:
\saveSpaceText
\noindent\textbf{\base}: \base is a vanilla baseline we designed, which uses the simplest matching method with basic features. It obtains a list of candidates based on common features that exist in both the comparison unit and detection target.  For each candidate, it will be reported as positive if: 1) the number of common features is larger than 15, or 2) the proportion of common features against candidate is greater than 0.2. Since the approach is simple and straightforward, \base method is very efficient.

\saveSpaceText
\noindent\textbf{\libdx}~\cite{libdx}: \libdx is  a tool for binary-to-binary comparison. It organizes constants as feature blocks to reduce false positives and take file names and requirement information as supplementary features. \libdx also relies on basic features.

\saveSpaceText
\noindent\textbf{\bsfinder} \cite{b2sfinder}: \bsfinder is a state-of-the-art tool in the area of binary-source TPL detection. It uses seven types of features (\verb|String|, \verb|Export|, \verb|Switch/case|, \verb|If/else|, \verb|String Array|, \verb|Integer Array|, \verb|Enum Array|) separately for detection. Results of each kind of features are simply aggregated together. For binary-to-binary comparison, only the first four types are applicable.

\section{Evaluation}\label{results}
In  our evaluation, we aim to answer the following research questions:
\subsection{RQ1: How does \ours perform in detecting TPLs in binaries?}
\label{sec:rq1}

We evaluate the TPL detection performance of \ours on the \datasetTwo by comparing it to the recent three baselines introduced in \Sec~\ref{sec:baseline_description}: \base, \libdx, and \bsfinder. We adopt Precision (P), Recall (R) and F1 score as evaluation metrics:
\begin{equation}
P = \frac{\#(correct \  libraries)}{\#(reported \  libraries)},\,\,\,\,R = \frac{\#(correct \ libraries)}{\#(libraries \  in \  target)}\nonumber
\end{equation}
\begin{equation}
F1 = \frac{2*P*R}{P+R}\nonumber
\end{equation}

Experimental results are shown in \Tab~\ref{tab:experiments}.
From the results in \Tab~\ref{tab:experiments}, we can see that \ours outperforms all the baselines in terms of F1 score. \base method and \libdx have comparable F1 scores, but \base shows better recall and \libdx has the highest precision. \bsfinder achieves the highest recall rate of 94.2\%.

After inspecting the results and comparing \ours with baselines, we find that rules of basic features are similar among all methods and most false positives are retrieved due to common basic features across libraries. For \ours, in basic feature channel before FCG filter, some similar functions exist in targets and candidates, especially standard library functions. They make our FCG filter recognize the false positive as a potential library reuse. Function vector channel uses contents of functions as features. It has the ability to report reused libraries that cannot reach the thresholds in rules with only few basic features. \libdx adopts feature block to eliminate false positives, however it significantly reduces the recall rate. \bsfinder has the lowest F1 score and precision, even though it has the highest recall. The new features (i.e., constants in ``switch/case'' and ``if/else'' statements) introduced by \bsfinder bring a slight improvement in recall, but more false positives. Besides, extra computation cost is caused by new features (\Sec~\ref{sec:rq5}).

\begin{table}[t]
  \caption{Performance comparison with baselines on the \datasetTwo.}
  \label{tab:experiments}
  \saveSpaceFig
  \begin{threeparttable}
    \begin{tabular}{lccccc}
    \toprule
    Method & P (\%) & R (\%)& F1& VD & VP (\%) \\\midrule
    \base & 29.1& 86.5& 0.44&1.32 &40.2\\
    \libdx & 88.4& 30.8& 0.46& /& /\\
    \bsfinder & 16.3& 94.2& 0.28 &/&/\\\midrule
    \ours & 55.3 & 93.6 & \textbf{0.70} & \textbf{0.58} & \textbf{60.4} \\
    \bottomrule
    \end{tabular}
  \end{threeparttable}
  \saveSpaceFig
\end{table}

\subsection{RQ2: How does \ours perform on the version identification task?}
\label{sec:rq4}

We use two metrics to evaluate version identification: VP (Version Precision) and VD (Version Distance). 
The VP metric calculates the rate of exact versions in all true positive libraries. It is used in previous works~\cite{Duan2017Identifying,zhang2019libid}.  
We design an additional metric VD to measure the \emph{distance} between the reported version and the correct version. 
Given two versions, $a1.b1.c1$, $a2.b2.c2$ in the format of Major.Minor.Patch, their distance is calculated by $10*\vert a1-a2\vert+b*\vert b1-b2\rvert+c*0.1*\lvert c1-c2\rvert$.
The reason for using VD is that vulnerabilities are usually associated with a series of consecutive versions. For example, \verb|CVE-2019-7317|\footnote{https://cve.mitre.org/cgi-bin/cvename.cgi?name=CVE-2019-7317} exists in \verb|LibPNG 1.6.x| ahead of \verb|1.6.37|. That means, if the reported version is not exactly the same as the ground truth, it still makes sense to report a version close to the true version. Besides, reporting a close version compensates for the cases when the true version is not in the TPL database.

For related work, \libdx and \bsfinder do not support version identification. Only OSSPolice proposes a simple version identification method, which uses unique features of one version across all versions of a library. We integrate this version identification method with \base as a baseline. From \Tab~\ref{tab:experiments}, we can see that \ours improves the version precision by 20\% and has a version distance of 0.58, which are much better than the \base combined with the unique feature based version identification method. The experimental results demonstrate that our fine-grained function features can effectively improve the performance of version identification in terms of both VD and VP.

\begin{table}[t]
  \caption{Subset1: function pairs with different operating systems. Subset2: different architectures. Subset3: different compilers. Subset4: different versions of the same compiler. Subset5: different optimization levels. Subset6: all five conditions are different, such as Linux-ARM-GCC-5.4-O3 and Win-x86-GCC-8.1-O0.}
  \label{tab:subsets}
  \saveSpaceFig
  \renewcommand\tabcolsep{2.7pt}
  \scalebox{0.9}{
  \begin{tabular}{lcccc}
    \toprule
    Subset&True Pairs& Recall (\%)\\
    \midrule
    Subset1 (OS) & 65,120 & 98.63\\
    Subset2 (ARM | x86)& 30,979& 96.99\\
    Subset3 (GCC | LLVM)& 69,783& 91.29\\
    Subset4 (Comp. Version)& 64,077& 98.17\\
    Subset5 (O0 | O3)& 106,420& 70.73\\
    Subset6 (Max Diff)& 5,690& 65.38\\
    \bottomrule
  \end{tabular}
  }
  \saveSpaceFig
\end{table}
\subsection{RQ3: How effective is the function retrieval component?}
\label{sec:rq2}

\subsubsection{Accuracy of function embedding}\label{accuracy of network model}
The performance of function retrieval module depends on the function embedding network. We first evaluate the accuracy of the function embedding network and determine the threshold for cosine similarity between two function embeddings. The threshold is set to 0.8 according to the validation dataset in the \datasetOne.

If a compiled instance changes a lot compared to another instance of the same function, they would be judged as dissimilar functions. Since we retrieve similar functions as features to find reused libraries, it is crucial to recall true similar pairs. We are concerned about whether \ours can tell that two functions are similar when the functions change due to compilation. In order to investigate the impact of different compilation conditions on code changes, we generate 8 subsets depending on the change conditions from test set. All datasets are described in Table~\ref{tab:subsets}. True pairs indicate all similar function pairs in that subset. Recall represents how many similar function pairs can be correctly judged when performing function comparison. We use the trained model to evaluate the impact of each condition on the recall of the similar functions, since recall is more important than precision in the retrieval phase.

In \Tab~\ref{tab:subsets}, We can see that compilation conditions, operating system, and compiler version, have less impact on binary code change. Even if code is changed by these two conditions, 98\%-99\% similar function pairs could still be correctly identified as similar pairs. Different architectures cause slightly greater impact on binary code and different compilers change the binary code more. The optimization level O0-O3 has the most significant impact on code changes and only 70\% similar function pairs can be recalled. As most conditions are different, The subset6 (maximum differences) only achieves a recall of 65\%.

\subsubsection{Similar function retrieval}

\begin{table}[t]
  \caption{Recall of function retrieval on the \datasetOne against the \datasetThree.}
  \label{tab:function retrieval}
  \saveSpaceFig
  \renewcommand\tabcolsep{2.5pt}
  \scalebox{0.9}{
  \begin{tabular}{lcccc}
    \toprule
    Subset&top-10 (\%) &top-20 (\%) & top-50 (\%)& top-100 (\%)\\
    \midrule
        Subset1-1 (Linux | Win)&59.13&63.43&67.66&71.07\\
        Subset1-2 (Linux | Mac)&73.60&75.82&78.22&79.77\\
        Subset1-3 (Win | Mac)&55.66&58.25&62.35&66.20\\
        Subset2 (ARM | x86)&18.76&24.66&29.65&33.61\\
        Subset3 (GCC | LLVM)&23.86&27.21&31.96&35.52\\
        Subset4 (Comp. Version)&70.53&74.28&77.49&79.72\\
        Subset5-1 (O0 | O3)&2.29&2.68&3.63&4.56\\
        Subset5-2 (O2 | O0)&3.41&4.21&5.55&6.84\\
        Subset5-3 (O2 | O1)&33.56&36.32&39.97&42.46\\
        Subset5-4 (O2 | O3)&65.31&66.38&67.76&68.76\\
        Subset6 (Max Diff)&0.54&0.76&1.09&1.48\\
  \bottomrule
\end{tabular}
}
\saveSpaceFig
\end{table}

\begin{table*}[t]
 \caption{Impacts of the function retrieval module on data with different proportions of basic features. ``+fr'' indicates that our function retrieval module is used, and ``-fr'' denotes that our function retrieval module is removed.}
 \saveSpaceFig
  \label{tab:fr_evaluation}
  \begin{tabular}{l ccc ccc ccc ccc}
    \toprule
    \multirow{2}{*}{String Percentage} & \multicolumn{3}{c}{0\%} & \multicolumn{3}{c}{25\%} & \multicolumn{3}{c}{50\%} & \multicolumn{3}{c}{100\%}\\
    \cmidrule(r){2-4}  \cmidrule(r){5-7} \cmidrule(r){8-10} \cmidrule(r){11-13}
    & P (\%) & R (\%)& F1 & P (\%) & R (\%)& F1 & P (\%) & R (\%)& F1 & P (\%) & R (\%)& F1 \\
    \midrule
    \base & / & 0 & / & 86.0 & 45.6 & 0.60 &74.1&77.1& 0.76 &29.1&86.5& 0.44\\
    \baseWithFR &96.5&32.2& 0.48 &84.9&57.9& 0.69 &73.6&83.0& 0.78 & 29.2&88.3& 0.44\\
    \midrule
    \bsfinder &6.5&17.5& 0.10 &19.6&71.9& 0.31 & 26.4&86.0&0.40&16.3&94.2& 0.28\\
    \bsfinderWithFR& 12.8& 35.7& 0.19 & 19.8 & 73.1 & 0.31 &26.4&86.5& 0.41 & 16.3 & 94.2& 0.28\\
    \midrule
    \oursWithoutFR& / &0& / & 63.7 &65.5 &0.65 & 69.8 & 82.5 & 0.76 & 55.7 & 91.8& 0.69\\
    \ours& 96.5& 32.2& 0.48 &65.5&72.5 & 0.69 & 70.3 & 86.0 & 0.77 & 55.3 & 93.6 & 0.70 \\
    \bottomrule
\end{tabular}
\end{table*}

Because we retrieve similar functions as clues to find potential reused libraries, it is critical to recall enough similar functions between detection target and reused library. Next, we evaluate the performance of function retrieval against the function vector database. Starting from similar function pairs of each subset, $\langle f_i, f_i^{'}\rangle$, we add $f_i$ into function vector database and search $f_i^{'}$ against the database. We consider $f_i$ to be recalled if it is in top-$K$ nearest neighbors and calculate the recall rate for each subset. In \Sec~\ref{accuracy of network model}, we know that different optimization levels (O0 | O3) result in quite different binary code. It may make the embedding model embed similar functions to vectors that have low cosine similarities. Thus, based on the \datasetOne, we generate more subsets with different optimization settings. Similarly, we further divide the subset1 into smaller subsets to explore the differences among Windows, Linux, and MacOS settings.

The recall values of different top-$K$ on all subsets are shown in \Tab~\ref{tab:function retrieval}. In general, recall has dropped a lot compared to similar function identification in \Sec~\ref{accuracy of network model}. As $K$ increases, the growth of recall becomes slower. A $K$ larger than 100 would result in little increase in recall, but retrieve more false positives.
In \Tab~\ref{tab:function retrieval}, subset1, subset4 and subset5-4 (O2 | O3) have higher recall values than the other subsets. Subset2, subset3 and subset5-3 (O2 | O1) have lower recall values. The subset5-1 (O0 | O3) and subset5-1 (O2 | O0) have the lowest recall values among all compilation conditions. Most optimization settings are disabled at O0 and the higher the optimization level is, the more optimizations are enabled.\footnote{https://gcc.gnu.org/onlinedocs/gcc/Optimize-Options.html}
Many compilation optimizations like \verb|-finline-functions-called-once| from O1, \verb|-finline-functions| from O2 and \verb|-fpeel-loops| at O3 can change the structure of CFG a lot and generate a different graph. It makes the function embedding network embed the function to a vector that has a small cosine similarity with similar functions compiled under other optimization levels.
We find that closer levels, such as (O2 | O1) and (O2, O3), have higher recall values.
As the most extreme case, the subset6 has the lowest recall value among all subsets. Even for the top-100 candidates, less than 2\% of similar functions are recalled. The problem of low recall on subset5-1, subset5-2, and subset6 is left to our future work. 

\begin{table}[t]
  \caption{Impacts of the  FCG filter component.}
  \saveSpaceFig
  \label{tab:fcg_evaluation}
  \begin{tabular}{lccc}
    \toprule
    Method & P (\%) & R (\%)& F1\\
    \midrule
    \oursWithoutFcg & 15.3 & 93.6 & 0.26 \\
    \ours & 55.3 & 93.6 & 0.70\\ \midrule
    \oursBaseWithoutFCG & 25.9& 91.8& 0.40\\
    \oursBase & 55.0 & 91.8 & 0.69\\ \midrule
    \oursFRWithoutFCG & 12.2 & 38.0 & 0.18\\
    \oursFR & 96.5 & 32.2 & 0.48\\ \midrule
    \base &29.1& 86.5& 0.43\\
    \baseWithFcg & 43.0& 79& 0.56\\ \midrule
    \bsfinder & 16.3& 94.2&0.28\\
    \bsfinderWithFcg & 41.4&94.2&0.58\\
    \bottomrule
\end{tabular}
\saveSpaceFig
\end{table}
\subsubsection{Effectiveness analysis of function retrieval module}
We analyze the effectiveness of the function retrieval module in improving the recall when no enough basic features are available. As described in \Sec~\ref{motivation}, most of string literals can be deleted by setting macros. Different levels of printouts (e.g., debug, warning, error) can be selected by developers. Therefore, different numbers of strings are kept in binaries. Exported function names, debloating techniques, and partial clones can delete unused functions in the library, leading to a reduction of the number of function name features. In order to simulate these scenarios, we randomly select basic features and keep 4 levels of proportions, 0\%, 25\%, 50\% and 100\%, and then conduct evaluations. Detection results on data with different proportions of basic features are provided in \Tab~\ref{tab:fr_evaluation}. It shows that our function retrieval module can bring different improvements to all methods. Improvements are more significant when there are fewer basic features. In extreme cases, for binaries without basic features, the function retrieval module can achieve a precision of 96.5\% and a recall of 32.2\% for \base. It is much better than \bsfinder that uses \verb|switch/else| and \verb|if/else|. We could find that when there is a sufficient quantity of basic features, libraries can be easily recalled. In the meantime, more false positives are reported, leading to a lower precision. From \Tab~\ref{tab:fr_evaluation}, we can see that the proportion of 50\% achieves the highest F1 than other proportions. That is because random selection eliminates some popular features, and the false positives caused by the popular features are reduced.
However, it does not mean that we can use only half of the basic features in detection. It cannot be assumed that there are always enough unique features in detection targets.

\subsection{RQ4: How effective is the FCG filter component?}
\label{sec:rq3}

We evaluate the effectiveness of FCG filter by adding it to baselines (i.e., variants \baseWithFcg and \bsfinderWithFcg) and removing it from \ours (i.e., variants \oursWithoutFcg, \oursBaseWithoutFCG and \oursFRWithoutFCG). \oursBase is \ours with only basic feature channel. \oursFR is \ours with only function vector channel. As shown in \Tab~\ref{tab:fcg_evaluation}, the use of FCG filter can significantly improve the precision of all evaluated approaches and their variants. 
Meanwhile, the recall rates generally remain the same. The recall rates of the baseline \oursFRWithoutFCG and \base dropped slightly because the FCG filter may filter out a small number of true positives. However, the F1 scores of all methods generally get improved. This study confirms the effectiveness of our FCG filter component for both \ours and other methods. Also, we find that common functions, especially standard library functions, cause many false positives. In our future work, we will explore the IDA FLIRT technique,\footnote{https://hex-rays.com/products/ida/tech/flirt/} which would be beneficial to recognize standard library functions generated by the supported compilers.

\subsection{RQ5: How efficient is \ours?}
\label{sec:rq5}

We use five servers to extract features. Each server has 8-core Intel i7 CPU and 64GB memory. For model training and evaluating \ours, the system is deployed on a server with a 20-core Intel E5 CPU, 256GB memory and 4 Nvidia Tesla V100 GPUs.
Total time consumption contains two parts: extracting features of TPLs from the \datasetThree and the detection time. 

It is one-time cost to process \datasetThree and extract features of TPLs, although it is time-consuming and resource-consuming. 
This process took us about 43 hours on five servers.
More than 95\% time spent on disassembling binary code, since reverse engineering is a time-consuming task, especially for large binaries. We set a timeout threshold to kill the task thread, if it cannot be finished within 30 minutes. According to our records, less than 1\% binaries in \datasetThree encountered timeout.
As for detection, \base method is most efficient and only takes 5 minutes. It uses the simplest matching method and can search features based on an inverted scheme. Compared to \base method, \libdx first performs the same feature searching task. Then it uses feature block method and takes 25 minutes to filter out false positives. 
Basic features in \bsfinder supports accelerating the matching process using an inverted index. But \verb|switch/case| and \verb|if/else| can only be applied by one-to-one comparison against \datasetThree. In total, \bsfinder requires 70 hours during the detection process.
For \ours, its detection time consists of three parts: extracting features from detection targets costs 35 minutes, basic feature channel costs 7.5 hours, and function vector channel costs 18 minutes. Basic feature channel requires more time, since there is a larger number of initial candidates to be passed into FCG filter component. FCG filter is the most time-consuming component in \ours. In Basic feature channel, basic features matching can be finished within several minutes, like \base method. The rest of time (more than 7 hours) is devoted to FCG filter. In function vector channel, function retrieval is working on Milvus. It takes 1.1 seconds to retrieve the top-$100$ most similar function per 1,000 queries.

\section{Threats to Validity}\label{threats}
There are three main threats to validity:
\saveSpaceText
\noindent\textbf{Binary obfuscation.}
Some libraries could utilize code obfuscation techniques to protect their intellectual property. The  obfuscated code often has poor readability and maintainability, and could significantly change the features \ours uses.
Currently, LibDB does not consider code obfuscation. We will handle the commonly-used 
code obfuscation techniques (such as string encryption and CFG-flattening) in our future work.

\saveSpaceText
\noindent\textbf{New library version.}
\ours detects TPLs based on a local database containing features of TPLs that are built in advance. It can only report possible reuses in the database. To mitigate this threat, we build a large-scale database containing 997 libraries with 25,000 version. The database could also be regularly updated with new libraries and new versions.
Still, if the detection target is not in our database, \ours will only report its closest version.

\saveSpaceText
\noindent\textbf{Establishment of TPL database.} 
Currently, there is no public TPL database for binaries. Therefore, we have to build a new database from Fedora mirrors. We also search the library names in the NVD database. If there are vulnerabilities related to the named library, we will keep the library in our TPL database. This is a relatively simple method to identify which libraries are related to each of the vulnerabilities in NVD. Automated identification of libraries from vulnerability data is a challenging problem~\cite{chen2020automated} and will be an important future work.

\section{Related work}\label{related work}

\saveSpaceText
\noindent\textbf{Third-party library detection.}
Most existing works for TPL detection are designed for Java libraries in Android applications~\cite{Zhan2021ResearchOT}. Java is a cross-platform language. Java bytecode compiled in different compilation scenarios remain unchanged. In addition to string literals, there are more features such as class dependency~\cite{zhang2019libid, Zhang2018DetectingTL}, CFG centroid~\cite{Duan2017Identifying}, semantic features from program dependency graphs~\cite{Crussell2015AnDarwinSD}, and opcode of CFG~\cite{Zhan2021ATVHunterRV} can be used to build fingerprints. Bytecode has different formats compared to binary code and these features is not effective for binaries. Therefore, techniques for Java library detection can not be directly  applied for TPL detection in binaries.
As discussed in \Sec~\ref{motivation}, there are also several work on TPL detection for binaries, including BAT~\cite{Hemel2011FindingSL}, OSSPolic~\cite{Duan2017Identifying}, \libdx~\cite{libdx},  and \bsfinder~\cite{b2sfinder}.

\saveSpaceText
\noindent\textbf{Binary code clone detection.} 
Existing binary code clone detection approaches mostly focus on function-level~\cite{Chandramohan2016BinGoCC, Xue2019AccurateAS} or basic-block-level~\cite{Pewny2015CrossArchitectureBS, Luo2014SemanticsbasedOB} comparison such as vulnerable function detection and bug search. These approaches are based on CFG comparison or instruction analysis which are very slow and not scalable when analyzing a large-scale code base. DiscovRE~\cite{Eschweiler2016discovREEC} employs a pre-filter based on numeric features to retrieve a small set of candidates. Genius~\cite{Feng2016ScalableGB} is the first method to embed CFGs and apply function vector search to detect similar binary code. Recently, the development of graph embedding techniques using neural network has inspired researchers to embed functions to vectors~\cite{Xu2017NeuralNG, Yu2020OrderMS, Gao2018VulSeekerAS} for binary code clone detection. It can efficiently retrieve similar functions from a large-scale database.

\saveSpaceText
\noindent\textbf{Similarity by composition.}  Inspired by image similarity~\cite{Boiman2006SimilarityBC}, Yaniv et al.~\cite{david2016statistical} illustrate the concept ``binary code similarity by composition'', which means that a binary file can be composed of parts of other binary files. It is similar to the concept of ``fused binary'' in our study. The difference is that Yaniv et al. consider instruction-level snippets and calculate the similarity of basic-block slices, while we focus on the similarity calculation at the file level.

\section{Conclusion}\label{conclusion}
In this paper, we propose a framework, \ours, for binary-oriented TPL detection. \ours uses contents of functions as features. It embeds functions in such a way that the similar functions have a higher cosine similarity score than the dissimilar functions. We obtain two lists of initial candidates via the basic feature channel and the function vector channel.
Candidates from two channels, after FCG filtering, are combined as the final detection results.
\ours is able to further provide version identification of TPLs contained in the detection target.
Extensive experiments have demonstrated the effectiveness and efficiency of \ours. 

Our datasets and source code are available at \textbf{\url{https://github.com/DeepSoftwareAnalytics/LibDB}}.

\begin{acks}
We thank the anonymous reviewers for their helpful feedback. 
\end{acks}

\balance
\bibliographystyle{ACM-Reference-Format}
\bibliography{ref}


\begin{thebibliography}{39}


\ifx \showCODEN    \undefined \def \showCODEN     #1{\unskip}     \fi
\ifx \showDOI      \undefined \def \showDOI       #1{#1}\fi
\ifx \showISBNx    \undefined \def \showISBNx     #1{\unskip}     \fi
\ifx \showISBNxiii \undefined \def \showISBNxiii  #1{\unskip}     \fi
\ifx \showISSN     \undefined \def \showISSN      #1{\unskip}     \fi
\ifx \showLCCN     \undefined \def \showLCCN      #1{\unskip}     \fi
\ifx \shownote     \undefined \def \shownote      #1{#1}          \fi
\ifx \showarticletitle \undefined \def \showarticletitle #1{#1}   \fi
\ifx \showURL      \undefined \def \showURL       {\relax}        \fi
\providecommand\bibfield[2]{#2}
\providecommand\bibinfo[2]{#2}
\providecommand\natexlab[1]{#1}
\providecommand\showeprint[2][]{arXiv:#2}

\bibitem[\protect\citeauthoryear{Agadakos, Jin, Williams-King, Kemerlis, and
  Portokalidis}{Agadakos et~al\mbox{.}}{2019}]%
        {Agadakos2019NibblerDB}
\bibfield{author}{\bibinfo{person}{Ioannis Agadakos}, \bibinfo{person}{Di Jin},
  \bibinfo{person}{David Williams-King}, \bibinfo{person}{V.~P. Kemerlis},
  {and} \bibinfo{person}{G. Portokalidis}.} \bibinfo{year}{2019}\natexlab{}.
\newblock \showarticletitle{Nibbler: debloating binary shared libraries}.
\newblock \bibinfo{journal}{\emph{Proceedings of the 35th Annual Computer
  Security Applications Conference}} (\bibinfo{year}{2019}).
\newblock


\bibitem[\protect\citeauthoryear{Boiman and Irani}{Boiman and Irani}{2006}]%
        {Boiman2006SimilarityBC}
\bibfield{author}{\bibinfo{person}{Oren Boiman} {and} \bibinfo{person}{Michal
  Irani}.} \bibinfo{year}{2006}\natexlab{}.
\newblock \showarticletitle{Similarity by Composition}. In
  \bibinfo{booktitle}{\emph{Advances in Neural Information Processing Systems
  (NIPS 2006)}}.
\newblock


\bibitem[\protect\citeauthoryear{Bromley, Guyon, LeCun, S{\"a}ckinger, and
  Shah}{Bromley et~al\mbox{.}}{1993}]%
        {bromley1993signature}
\bibfield{author}{\bibinfo{person}{Jane Bromley}, \bibinfo{person}{Isabelle
  Guyon}, \bibinfo{person}{Yann LeCun}, \bibinfo{person}{Eduard S{\"a}ckinger},
  {and} \bibinfo{person}{Roopak Shah}.} \bibinfo{year}{1993}\natexlab{}.
\newblock \showarticletitle{Signature verification using a" siamese" time delay
  neural network}.
\newblock \bibinfo{journal}{\emph{Advances in neural information processing
  systems}}  \bibinfo{volume}{6} (\bibinfo{year}{1993}),
  \bibinfo{pages}{737--744}.
\newblock


\bibitem[\protect\citeauthoryear{Chandramohan, Xue, Xu, Liu, Cho, and
  Tan}{Chandramohan et~al\mbox{.}}{2016}]%
        {Chandramohan2016BinGoCC}
\bibfield{author}{\bibinfo{person}{Mahinthan Chandramohan},
  \bibinfo{person}{Yinxing Xue}, \bibinfo{person}{Zhengzi Xu},
  \bibinfo{person}{Yang Liu}, \bibinfo{person}{Chia~Yuan Cho}, {and}
  \bibinfo{person}{Hee Beng~Kuan Tan}.} \bibinfo{year}{2016}\natexlab{}.
\newblock \showarticletitle{BinGo: cross-architecture cross-OS binary search}.
\newblock \bibinfo{journal}{\emph{Proceedings of the 2016 24th ACM SIGSOFT
  International Symposium on Foundations of Software Engineering}}
  (\bibinfo{year}{2016}).
\newblock


\bibitem[\protect\citeauthoryear{Chen, Santosa, Sharma, and Lo}{Chen
  et~al\mbox{.}}{2020}]%
        {chen2020automated}
\bibfield{author}{\bibinfo{person}{Yang Chen}, \bibinfo{person}{Andrew~E
  Santosa}, \bibinfo{person}{Asankhaya Sharma}, {and} \bibinfo{person}{David
  Lo}.} \bibinfo{year}{2020}\natexlab{}.
\newblock \showarticletitle{Automated identification of libraries from
  vulnerability data}. In \bibinfo{booktitle}{\emph{Proceedings of the ACM/IEEE
  42nd International Conference on Software Engineering: Software Engineering
  in Practice}}. \bibinfo{pages}{90--99}.
\newblock


\bibitem[\protect\citeauthoryear{Crussell, Gibler, and Chen}{Crussell
  et~al\mbox{.}}{2015}]%
        {Crussell2015AnDarwinSD}
\bibfield{author}{\bibinfo{person}{J. Crussell}, \bibinfo{person}{Clint
  Gibler}, {and} \bibinfo{person}{Hao Chen}.} \bibinfo{year}{2015}\natexlab{}.
\newblock \showarticletitle{AnDarwin: Scalable Detection of Android Application
  Clones Based on Semantics}.
\newblock \bibinfo{journal}{\emph{IEEE Transactions on Mobile Computing}}
  \bibinfo{volume}{14} (\bibinfo{year}{2015}), \bibinfo{pages}{2007--2019}.
\newblock


\bibitem[\protect\citeauthoryear{Dai, Dai, and Song}{Dai et~al\mbox{.}}{2016}]%
        {dai2016discriminative}
\bibfield{author}{\bibinfo{person}{Hanjun Dai}, \bibinfo{person}{Bo Dai}, {and}
  \bibinfo{person}{Le Song}.} \bibinfo{year}{2016}\natexlab{}.
\newblock \showarticletitle{Discriminative embeddings of latent variable models
  for structured data}. In \bibinfo{booktitle}{\emph{International conference
  on machine learning}}. PMLR, \bibinfo{pages}{2702--2711}.
\newblock


\bibitem[\protect\citeauthoryear{David, Partush, and Yahav}{David
  et~al\mbox{.}}{2016}]%
        {david2016statistical}
\bibfield{author}{\bibinfo{person}{Yaniv David}, \bibinfo{person}{Nimrod
  Partush}, {and} \bibinfo{person}{Eran Yahav}.}
  \bibinfo{year}{2016}\natexlab{}.
\newblock \showarticletitle{Statistical similarity of binaries}.
\newblock \bibinfo{journal}{\emph{ACM SIGPLAN Notices}} \bibinfo{volume}{51},
  \bibinfo{number}{6} (\bibinfo{year}{2016}), \bibinfo{pages}{266--280}.
\newblock


\bibitem[\protect\citeauthoryear{Ding, Fung, and Charland}{Ding
  et~al\mbox{.}}{2019}]%
        {ding2019asm2vec}
\bibfield{author}{\bibinfo{person}{Steven~HH Ding},
  \bibinfo{person}{Benjamin~CM Fung}, {and} \bibinfo{person}{Philippe
  Charland}.} \bibinfo{year}{2019}\natexlab{}.
\newblock \showarticletitle{Asm2vec: Boosting static representation robustness
  for binary clone search against code obfuscation and compiler optimization}.
  In \bibinfo{booktitle}{\emph{2019 IEEE Symposium on Security and Privacy
  (SP)}}. IEEE, \bibinfo{pages}{472--489}.
\newblock


\bibitem[\protect\citeauthoryear{Duan, Bijlani, Xu, Kim, and Lee}{Duan
  et~al\mbox{.}}{2017}]%
        {Duan2017Identifying}
\bibfield{author}{\bibinfo{person}{Ruian Duan}, \bibinfo{person}{Ashish
  Bijlani}, \bibinfo{person}{Meng Xu}, \bibinfo{person}{Taesoo Kim}, {and}
  \bibinfo{person}{Wenke Lee}.} \bibinfo{year}{2017}\natexlab{}.
\newblock \showarticletitle{Identifying Open-Source License Violation and 1-day
  Security Risk at Large Scale}. In \bibinfo{booktitle}{\emph{ACM Sigsac
  Conference}}. \bibinfo{pages}{2169--2185}.
\newblock


\bibitem[\protect\citeauthoryear{Durumeric, Li, Kasten, Amann, Beekman, Payer,
  Weaver, Adrian, Paxson, Bailey, et~al\mbox{.}}{Durumeric
  et~al\mbox{.}}{2014}]%
        {durumeric2014matter}
\bibfield{author}{\bibinfo{person}{Zakir Durumeric}, \bibinfo{person}{Frank
  Li}, \bibinfo{person}{James Kasten}, \bibinfo{person}{Johanna Amann},
  \bibinfo{person}{Jethro Beekman}, \bibinfo{person}{Mathias Payer},
  \bibinfo{person}{Nicolas Weaver}, \bibinfo{person}{David Adrian},
  \bibinfo{person}{Vern Paxson}, \bibinfo{person}{Michael Bailey},
  {et~al\mbox{.}}} \bibinfo{year}{2014}\natexlab{}.
\newblock \showarticletitle{The matter of heartbleed}. In
  \bibinfo{booktitle}{\emph{Proceedings of the 2014 conference on internet
  measurement conference}}. \bibinfo{pages}{475--488}.
\newblock


\bibitem[\protect\citeauthoryear{Eschweiler, Yakdan, and
  Gerhards-Padilla}{Eschweiler et~al\mbox{.}}{2016}]%
        {Eschweiler2016discovREEC}
\bibfield{author}{\bibinfo{person}{Sebastian Eschweiler},
  \bibinfo{person}{Khaled Yakdan}, {and} \bibinfo{person}{E.
  Gerhards-Padilla}.} \bibinfo{year}{2016}\natexlab{}.
\newblock \showarticletitle{discovRE: Efficient Cross-Architecture
  Identification of Bugs in Binary Code}. In \bibinfo{booktitle}{\emph{NDSS}}.
\newblock


\bibitem[\protect\citeauthoryear{Feng, Yuan, Li, Ban, Xiao, Wang, Tang, Su, Yu,
  Xu, Piao, Xue, and Huo}{Feng et~al\mbox{.}}{2019}]%
        {b2sfinder}
\bibfield{author}{\bibinfo{person}{Muyue Feng}, \bibinfo{person}{Zimu Yuan},
  \bibinfo{person}{Feng Li}, \bibinfo{person}{Gu Ban}, \bibinfo{person}{Yang
  Xiao}, \bibinfo{person}{Shiyang Wang}, \bibinfo{person}{Qian Tang},
  \bibinfo{person}{He Su}, \bibinfo{person}{Chendong Yu},
  \bibinfo{person}{Jiahuan Xu}, \bibinfo{person}{Aihua Piao},
  \bibinfo{person}{Jingling Xue}, {and} \bibinfo{person}{Wei Huo}.}
  \bibinfo{year}{2019}\natexlab{}.
\newblock \showarticletitle{B2SFinder: Detecting Open-Source Software Reuse in
  COTS Software}. In \bibinfo{booktitle}{\emph{Proceedings of the 34th IEEE/ACM
  International Conference on Automated Software Engineering}} (San Diego,
  California) \emph{(\bibinfo{series}{ASE '19})}. \bibinfo{publisher}{IEEE
  Press}, \bibinfo{pages}{1038–1049}.
\newblock
\showISBNx{9781728125084}
\urldef\tempurl%
\url{https://doi.org/10.1109/ASE.2019.00100}
\showDOI{\tempurl}


\bibitem[\protect\citeauthoryear{Feng, Zhou, Xu, Cheng, Testa, and Yin}{Feng
  et~al\mbox{.}}{2016a}]%
        {genius}
\bibfield{author}{\bibinfo{person}{Qian Feng}, \bibinfo{person}{Rundong Zhou},
  \bibinfo{person}{Chengcheng Xu}, \bibinfo{person}{Yao Cheng},
  \bibinfo{person}{Brian Testa}, {and} \bibinfo{person}{Heng Yin}.}
  \bibinfo{year}{2016}\natexlab{a}.
\newblock \showarticletitle{Scalable Graph-Based Bug Search for Firmware
  Images}. In \bibinfo{booktitle}{\emph{Proceedings of the 2016 ACM SIGSAC
  Conference on Computer and Communications Security}} (Vienna, Austria)
  \emph{(\bibinfo{series}{CCS '16})}. \bibinfo{publisher}{Association for
  Computing Machinery}, \bibinfo{address}{New York, NY, USA},
  \bibinfo{pages}{480–491}.
\newblock
\showISBNx{9781450341394}
\urldef\tempurl%
\url{https://doi.org/10.1145/2976749.2978370}
\showDOI{\tempurl}


\bibitem[\protect\citeauthoryear{Feng, Zhou, Xu, Cheng, Testa, and Yin}{Feng
  et~al\mbox{.}}{2016b}]%
        {Feng2016ScalableGB}
\bibfield{author}{\bibinfo{person}{Qian Feng}, \bibinfo{person}{Rundong Zhou},
  \bibinfo{person}{Chengcheng Xu}, \bibinfo{person}{Yao Cheng},
  \bibinfo{person}{Brian Testa}, {and} \bibinfo{person}{Heng Yin}.}
  \bibinfo{year}{2016}\natexlab{b}.
\newblock \showarticletitle{Scalable Graph-based Bug Search for Firmware
  Images}.
\newblock \bibinfo{journal}{\emph{Proceedings of the 2016 ACM SIGSAC Conference
  on Computer and Communications Security}} (\bibinfo{year}{2016}).
\newblock


\bibitem[\protect\citeauthoryear{Gao, Yang, Fu, Jiang, and Sun}{Gao
  et~al\mbox{.}}{2018}]%
        {Gao2018VulSeekerAS}
\bibfield{author}{\bibinfo{person}{J. Gao}, \bibinfo{person}{X. Yang},
  \bibinfo{person}{Ying Fu}, \bibinfo{person}{Yu Jiang}, {and}
  \bibinfo{person}{Jia-Guang Sun}.} \bibinfo{year}{2018}\natexlab{}.
\newblock \showarticletitle{VulSeeker: A Semantic Learning Based Vulnerability
  Seeker for Cross-Platform Binary}.
\newblock \bibinfo{journal}{\emph{2018 33rd IEEE/ACM International Conference
  on Automated Software Engineering (ASE)}} (\bibinfo{year}{2018}),
  \bibinfo{pages}{896--899}.
\newblock


\bibitem[\protect\citeauthoryear{GNU}{GNU}{2021a}]%
        {gcc}
\bibfield{author}{\bibinfo{person}{GNU}.} \bibinfo{year}{2021}\natexlab{a}.
\newblock \bibinfo{title}{GCC, the GNU Compiler Collection}.
\newblock
\newblock
\urldef\tempurl%
\url{https://gcc.gnu.org/}
\showURL{%
Retrieved Sep 9, 2021 from \tempurl}


\bibitem[\protect\citeauthoryear{GNU}{GNU}{2021b}]%
        {llvm}
\bibfield{author}{\bibinfo{person}{GNU}.} \bibinfo{year}{2021}\natexlab{b}.
\newblock \bibinfo{title}{The LLVM Compiler Infrastructure}.
\newblock
\newblock
\urldef\tempurl%
\url{https://llvm.org/}
\showURL{%
Retrieved May 9, 2021 from \tempurl}


\bibitem[\protect\citeauthoryear{Hemel, Kalleberg, Vermaas, and Dolstra}{Hemel
  et~al\mbox{.}}{2011}]%
        {Hemel2011FindingSL}
\bibfield{author}{\bibinfo{person}{Armijn Hemel}, \bibinfo{person}{K.~T.
  Kalleberg}, \bibinfo{person}{R. Vermaas}, {and} \bibinfo{person}{E.
  Dolstra}.} \bibinfo{year}{2011}\natexlab{}.
\newblock \showarticletitle{Finding software license violations through binary
  code clone detection}. In \bibinfo{booktitle}{\emph{MSR '11}}.
\newblock


\bibitem[\protect\citeauthoryear{Jang, Agrawal, and Brumley}{Jang
  et~al\mbox{.}}{2012}]%
        {Jang2012ReDeBugFU}
\bibfield{author}{\bibinfo{person}{Jiyong Jang}, \bibinfo{person}{Abeer
  Agrawal}, {and} \bibinfo{person}{David Brumley}.}
  \bibinfo{year}{2012}\natexlab{}.
\newblock \showarticletitle{ReDeBug: Finding Unpatched Code Clones in Entire OS
  Distributions}.
\newblock \bibinfo{journal}{\emph{2012 IEEE Symposium on Security and Privacy}}
  (\bibinfo{year}{2012}), \bibinfo{pages}{48--62}.
\newblock


\bibitem[\protect\citeauthoryear{Kim, Woo, Lee, and Oh}{Kim
  et~al\mbox{.}}{2017}]%
        {Kim2017VUDDYAS}
\bibfield{author}{\bibinfo{person}{Seulbae Kim}, \bibinfo{person}{Seunghoon
  Woo}, \bibinfo{person}{Heejo Lee}, {and} \bibinfo{person}{Hakjoo Oh}.}
  \bibinfo{year}{2017}\natexlab{}.
\newblock \showarticletitle{VUDDY: A Scalable Approach for Vulnerable Code
  Clone Discovery}.
\newblock \bibinfo{journal}{\emph{2017 IEEE Symposium on Security and Privacy
  (SP)}} (\bibinfo{year}{2017}), \bibinfo{pages}{595--614}.
\newblock


\bibitem[\protect\citeauthoryear{Li, Wang, Wang, Wang, Wu, Liu, Xue, and
  Huo}{Li et~al\mbox{.}}{2017}]%
        {li2017libd}
\bibfield{author}{\bibinfo{person}{Menghao Li}, \bibinfo{person}{Wei Wang},
  \bibinfo{person}{Pei Wang}, \bibinfo{person}{Shuai Wang},
  \bibinfo{person}{Dinghao Wu}, \bibinfo{person}{Jian Liu},
  \bibinfo{person}{Rui Xue}, {and} \bibinfo{person}{Wei Huo}.}
  \bibinfo{year}{2017}\natexlab{}.
\newblock \showarticletitle{Libd: Scalable and precise third-party library
  detection in android markets}. In \bibinfo{booktitle}{\emph{2017 IEEE/ACM
  39th International Conference on Software Engineering (ICSE)}}. IEEE,
  \bibinfo{pages}{335--346}.
\newblock


\bibitem[\protect\citeauthoryear{Luo, Ming, Wu, Liu, and Zhu}{Luo
  et~al\mbox{.}}{2014}]%
        {Luo2014SemanticsbasedOB}
\bibfield{author}{\bibinfo{person}{Lannan Luo}, \bibinfo{person}{Jiang Ming},
  \bibinfo{person}{Dinghao Wu}, \bibinfo{person}{Peng Liu}, {and}
  \bibinfo{person}{Sencun Zhu}.} \bibinfo{year}{2014}\natexlab{}.
\newblock \showarticletitle{Semantics-based obfuscation-resilient binary code
  similarity comparison with applications to software plagiarism detection}.
\newblock \bibinfo{journal}{\emph{Proceedings of the 22nd ACM SIGSOFT
  International Symposium on Foundations of Software Engineering}}
  (\bibinfo{year}{2014}).
\newblock


\bibitem[\protect\citeauthoryear{Pewny, Garmany, Gawlik, Rossow, and
  Holz}{Pewny et~al\mbox{.}}{2015}]%
        {Pewny2015CrossArchitectureBS}
\bibfield{author}{\bibinfo{person}{Jannik Pewny}, \bibinfo{person}{Behrad
  Garmany}, \bibinfo{person}{R. Gawlik}, \bibinfo{person}{C. Rossow}, {and}
  \bibinfo{person}{T. Holz}.} \bibinfo{year}{2015}\natexlab{}.
\newblock \showarticletitle{Cross-Architecture Bug Search in Binary
  Executables}.
\newblock \bibinfo{journal}{\emph{2015 IEEE Symposium on Security and Privacy}}
  (\bibinfo{year}{2015}), \bibinfo{pages}{709--724}.
\newblock


\bibitem[\protect\citeauthoryear{Quach, Prakash, and Yan}{Quach
  et~al\mbox{.}}{2018}]%
        {Quach2018DebloatingST}
\bibfield{author}{\bibinfo{person}{Anh Quach}, \bibinfo{person}{Aravind
  Prakash}, {and} \bibinfo{person}{Lok Yan}.} \bibinfo{year}{2018}\natexlab{}.
\newblock \showarticletitle{Debloating Software through Piece-Wise Compilation
  and Loading}. In \bibinfo{booktitle}{\emph{27th {USENIX} Security Symposium
  ({USENIX} Security 18)}}. \bibinfo{publisher}{{USENIX} Association},
  \bibinfo{address}{Baltimore, MD}, \bibinfo{pages}{869--886}.
\newblock
\showISBNx{978-1-939133-04-5}
\urldef\tempurl%
\url{https://www.usenix.org/conference/usenixsecurity18/presentation/quach}
\showURL{%
\tempurl}


\bibitem[\protect\citeauthoryear{synopsys}{synopsys}{2021}]%
        {blackduck}
\bibfield{author}{\bibinfo{person}{synopsys}.} \bibinfo{year}{2021}\natexlab{}.
\newblock \bibinfo{title}{Black Duck Software Composition Analysis}.
\newblock
\newblock
\urldef\tempurl%
\url{https://www.synopsys.com/software-integrity/security-testing/software-composition-analysis.html}
\showURL{%
Retrieved May 9, 2021 from \tempurl}


\bibitem[\protect\citeauthoryear{Tang, Luo, Fu, and Zhang}{Tang
  et~al\mbox{.}}{2020}]%
        {libdx}
\bibfield{author}{\bibinfo{person}{Wei Tang}, \bibinfo{person}{Ping Luo},
  \bibinfo{person}{Jialiang Fu}, {and} \bibinfo{person}{Dan Zhang}.}
  \bibinfo{year}{2020}\natexlab{}.
\newblock \showarticletitle{LibDX: A Cross-Platform and Accurate System to
  Detect Third-party Libraries in Binary code}. In
  \bibinfo{booktitle}{\emph{2019 IEEE 27th International Conference on Software
  Analysis, Evolution and Reengineering (SANER)}}. IEEE.
\newblock


\bibitem[\protect\citeauthoryear{Wang, Guo, Ma, and Chen}{Wang
  et~al\mbox{.}}{2015}]%
        {Wang2015WuKongAS}
\bibfield{author}{\bibinfo{person}{Haoyu Wang}, \bibinfo{person}{Yao Guo},
  \bibinfo{person}{Ziang Ma}, {and} \bibinfo{person}{X. Chen}.}
  \bibinfo{year}{2015}\natexlab{}.
\newblock \showarticletitle{WuKong: a scalable and accurate two-phase approach
  to Android app clone detection}.
\newblock \bibinfo{journal}{\emph{Proceedings of the 2015 International
  Symposium on Software Testing and Analysis}} (\bibinfo{year}{2015}).
\newblock


\bibitem[\protect\citeauthoryear{whitesource}{whitesource}{2021}]%
        {whitesource}
\bibfield{author}{\bibinfo{person}{whitesource}.}
  \bibinfo{year}{2021}\natexlab{}.
\newblock \bibinfo{title}{Automate your open source security and compliance
  workflows}.
\newblock
\newblock
\urldef\tempurl%
\url{https://www.whitesourcesoftware.com/}
\showURL{%
Retrieved May 9, 2021 from \tempurl}


\bibitem[\protect\citeauthoryear{Woo, Park, Kim, Lee, and Oh}{Woo
  et~al\mbox{.}}{2021}]%
        {woo2021centris}
\bibfield{author}{\bibinfo{person}{Seunghoon Woo}, \bibinfo{person}{Sunghan
  Park}, \bibinfo{person}{Seulbae Kim}, \bibinfo{person}{Heejo Lee}, {and}
  \bibinfo{person}{Hakjoo Oh}.} \bibinfo{year}{2021}\natexlab{}.
\newblock \showarticletitle{CENTRIS: A Precise and Scalable Approach for
  Identifying Modified Open-Source Software Reuse}. In
  \bibinfo{booktitle}{\emph{2021 IEEE/ACM 43rd International Conference on
  Software Engineering (ICSE)}}. IEEE, \bibinfo{pages}{860--872}.
\newblock


\bibitem[\protect\citeauthoryear{Xu, Liu, Feng, Yin, Song, and Song}{Xu
  et~al\mbox{.}}{2017}]%
        {Xu2017NeuralNG}
\bibfield{author}{\bibinfo{person}{Xiaojun Xu}, \bibinfo{person}{Chang Liu},
  \bibinfo{person}{Qian Feng}, \bibinfo{person}{Heng Yin}, \bibinfo{person}{Le
  Song}, {and} \bibinfo{person}{D. Song}.} \bibinfo{year}{2017}\natexlab{}.
\newblock \showarticletitle{Neural Network-based Graph Embedding for
  Cross-Platform Binary Code Similarity Detection}.
\newblock \bibinfo{journal}{\emph{Proceedings of the 2017 ACM SIGSAC Conference
  on Computer and Communications Security}} (\bibinfo{year}{2017}).
\newblock


\bibitem[\protect\citeauthoryear{Xue, Xu, Chandramohan, and Liu}{Xue
  et~al\mbox{.}}{2019}]%
        {Xue2019AccurateAS}
\bibfield{author}{\bibinfo{person}{Yinxing Xue}, \bibinfo{person}{Zhengzi Xu},
  \bibinfo{person}{Mahinthan Chandramohan}, {and} \bibinfo{person}{Y. Liu}.}
  \bibinfo{year}{2019}\natexlab{}.
\newblock \showarticletitle{Accurate and Scalable Cross-Architecture Cross-OS
  Binary Code Search with Emulation}.
\newblock \bibinfo{journal}{\emph{IEEE Transactions on Software Engineering}}
  \bibinfo{volume}{45} (\bibinfo{year}{2019}), \bibinfo{pages}{1125--1149}.
\newblock


\bibitem[\protect\citeauthoryear{Yu, Cao, Tang, Nie, Huang, and Wu}{Yu
  et~al\mbox{.}}{2020a}]%
        {Yu2020OrderMS}
\bibfield{author}{\bibinfo{person}{Zeping Yu}, \bibinfo{person}{Rui Cao},
  \bibinfo{person}{Qiyi Tang}, \bibinfo{person}{Sen Nie},
  \bibinfo{person}{Junzhou Huang}, {and} \bibinfo{person}{Shi Wu}.}
  \bibinfo{year}{2020}\natexlab{a}.
\newblock \showarticletitle{Order Matters: Semantic-Aware Neural Networks for
  Binary Code Similarity Detection}. In \bibinfo{booktitle}{\emph{AAAI}}.
\newblock


\bibitem[\protect\citeauthoryear{Yu, Zheng, Wang, Tang, Nie, and Wu}{Yu
  et~al\mbox{.}}{2020b}]%
        {Yu2020CodeCMRCR}
\bibfield{author}{\bibinfo{person}{Zeping Yu}, \bibinfo{person}{Wenxin Zheng},
  \bibinfo{person}{Jiaqi Wang}, \bibinfo{person}{Qiyi Tang},
  \bibinfo{person}{Sen Nie}, {and} \bibinfo{person}{Shi Wu}.}
  \bibinfo{year}{2020}\natexlab{b}.
\newblock \showarticletitle{CodeCMR: Cross-Modal Retrieval For Function-Level
  Binary Source Code Matching}. In \bibinfo{booktitle}{\emph{NeurIPS}}.
\newblock


\bibitem[\protect\citeauthoryear{Zhan, Fan, Chen, Wu, Liu, Luo, and Liu}{Zhan
  et~al\mbox{.}}{2021a}]%
        {Zhan2021ATVHunterRV}
\bibfield{author}{\bibinfo{person}{Xian Zhan}, \bibinfo{person}{Lingling Fan},
  \bibinfo{person}{Sen Chen}, \bibinfo{person}{Feng Wu},
  \bibinfo{person}{Tianming Liu}, \bibinfo{person}{Xiapu Luo}, {and}
  \bibinfo{person}{Yang Liu}.} \bibinfo{year}{2021}\natexlab{a}.
\newblock \showarticletitle{ATVHunter: Reliable Version Detection of
  Third-Party Libraries for Vulnerability Identification in Android
  Applications}.
\newblock \bibinfo{journal}{\emph{2021 IEEE/ACM 43rd International Conference
  on Software Engineering (ICSE)}} (\bibinfo{year}{2021}),
  \bibinfo{pages}{1695--1707}.
\newblock


\bibitem[\protect\citeauthoryear{Zhan, Fan, Liu, Chen, Li, Wang, Xu, Luo, and
  Liu}{Zhan et~al\mbox{.}}{2020}]%
        {zhan2020automated}
\bibfield{author}{\bibinfo{person}{Xian Zhan}, \bibinfo{person}{Lingling Fan},
  \bibinfo{person}{Tianming Liu}, \bibinfo{person}{Sen Chen},
  \bibinfo{person}{Li Li}, \bibinfo{person}{Haoyu Wang}, \bibinfo{person}{Yifei
  Xu}, \bibinfo{person}{Xiapu Luo}, {and} \bibinfo{person}{Yang Liu}.}
  \bibinfo{year}{2020}\natexlab{}.
\newblock \showarticletitle{Automated third-party library detection for android
  applications: Are we there yet?}. In \bibinfo{booktitle}{\emph{2020 35th
  IEEE/ACM International Conference on Automated Software Engineering (ASE)}}.
  IEEE, \bibinfo{pages}{919--930}.
\newblock


\bibitem[\protect\citeauthoryear{Zhan, Liu, Fan, Li, Chen, Luo, and Liu}{Zhan
  et~al\mbox{.}}{2021b}]%
        {Zhan2021ResearchOT}
\bibfield{author}{\bibinfo{person}{Xian Zhan}, \bibinfo{person}{Tianming Liu},
  \bibinfo{person}{Lingling Fan}, \bibinfo{person}{Li Li}, \bibinfo{person}{Sen
  Chen}, \bibinfo{person}{Xiapu Luo}, {and} \bibinfo{person}{Yang Liu}.}
  \bibinfo{year}{2021}\natexlab{b}.
\newblock \showarticletitle{Research on Third-Party Libraries in Android Apps:
  A Taxonomy and Systematic Literature Review}.
\newblock \bibinfo{journal}{\emph{IEEE Transactions on Software Engineering}}
  (\bibinfo{year}{2021}).
\newblock


\bibitem[\protect\citeauthoryear{Zhang, Beresford, and Kollmann}{Zhang
  et~al\mbox{.}}{2019}]%
        {zhang2019libid}
\bibfield{author}{\bibinfo{person}{Jiexin Zhang}, \bibinfo{person}{A.
  Beresford}, {and} \bibinfo{person}{Stephan~A. Kollmann}.}
  \bibinfo{year}{2019}\natexlab{}.
\newblock \showarticletitle{LibID: reliable identification of obfuscated
  third-party Android libraries}.
\newblock \bibinfo{journal}{\emph{Proceedings of the 28th ACM SIGSOFT
  International Symposium on Software Testing and Analysis}}
  (\bibinfo{year}{2019}).
\newblock


\bibitem[\protect\citeauthoryear{Zhang, Dai, Zhang, Huang, Yang, Yang, and
  Chen}{Zhang et~al\mbox{.}}{2018}]%
        {Zhang2018DetectingTL}
\bibfield{author}{\bibinfo{person}{Yuan Zhang}, \bibinfo{person}{Jiarun Dai},
  \bibinfo{person}{Xiaohan Zhang}, \bibinfo{person}{S. Huang},
  \bibinfo{person}{Zhemin Yang}, \bibinfo{person}{Min Yang}, {and}
  \bibinfo{person}{Hao Chen}.} \bibinfo{year}{2018}\natexlab{}.
\newblock \showarticletitle{Detecting third-party libraries in Android
  applications with high precision and recall}.
\newblock \bibinfo{journal}{\emph{2018 IEEE 25th International Conference on
  Software Analysis, Evolution and Reengineering (SANER)}}
  (\bibinfo{year}{2018}), \bibinfo{pages}{141--152}.
\newblock


\end{thebibliography}
\balance

\end{document}